\documentclass[prl,aps,twocolumn,showpacs,superscriptaddress]{revtex4}
\usepackage{amssymb}
\usepackage{amsmath}
\usepackage{graphicx}
\usepackage{wasysym}
\usepackage{color}

\usepackage{epstopdf}

\newcommand{\be}{\begin{eqnarray}}
\newcommand{\ee}{\end{eqnarray}}
\newcommand{\bbm}{\begin{bmatrix}}
\newcommand{\ebm}{\end{bmatrix}}
\newcommand{\bpm}{\begin{pmatrix}}
\newcommand{\epm}{\end{pmatrix}}
\renewcommand{\v}[1]{{\bf #1}}
\newcommand{\nn}{\nonumber \\}

\begin{document}
\title{Landau Level Quantization and Almost Flat Modes in Three-dimensional Semi-metals with Nodal Ring Spectra}

\author{Jun-Won \surname{Rhim}}
\affiliation{School of Physics, Korea Institute for Advanced Study, Seoul 130-722, Korea}

\author{Yong Baek \surname{Kim}}
\affiliation{Department of Physics,
University of Toronto, Toronto, Ontario M5S 1A7, Canada}
\affiliation{Canadian Institute for Advanced Research, Toronto, Ontario, M5G 1Z8, Canada}


\date{\today}

\begin{abstract}
We investigate novel Landau level structures of semi-metals with nodal ring dispersions. When the magnetic field is applied parallel to the plane in which the ring lies, there exist almost non-dispersive Landau levels at the Fermi level ($E_F=0$) as a function of the momentum along the field direction inside the ring. We show that the Landau levels at each momentum along the field direction can be described by the Hamiltonian for the graphene bilayer with fictitious inter-layer couplings under a tilted magnetic field. Near the center of the ring where the inter-layer coupling is negligible, we have Dirac Landau levels which explain the appearance of the zero modes. Although the inter-layer hopping amplitudes become finite at higher momenta, the splitting of zero modes is exponentially small and they remain almost flat due to the finite artificial in-plane component of the magnetic field. The emergence of the density of states peak at the Fermi level would be a hallmark of the ring dispersion.
\end{abstract}



\maketitle


\textit{Introduction.}--- Semi-metals, usually the reflection of unconventional electronic structures at the Fermi surface (FS), are related to various anomalous properties and/or exotic phases such as unconventional quantum Hall effect (QHE) in graphene systems \cite{novoselov,zhang1,castro,mccann}, pressure induced anomalous Hall effect in the Weyl semi-metal (SM) \cite{yang} and non-Fermi liquid phase and peculiar quantum oscillations in the quadratic band-touching SM \cite{yang2,krempa,moon,qi,rhim}.
Also, in many cases, they are classified as topologically non-trivial metals involving surface states \cite{burkov,wan,xu,young,wang} which are generalizations of the concept of the topological insulator \cite{hasan} to the metallic systems \cite{matsuura}.

Recently, there have been many suggestions for the novel semi-metals with nodal ring FS with different topological classification schemes \cite{carter,chen,kim,xie,zeng,xu2,zeng1,zeng2,zeng3}.
Their topological non-triviality ensures the existence of surface modes protected by inversion, time-reversal or certain lattice symmetries.
Since those candidate  materials for the nodal ring semi-metal (NRS) are proposed very recently, the investigations of their physical properties and experimental observations are still outstanding open problems.

In this work, we demonstrate that the NRSs exhibit unusual 3D Landau level structures when the magnetic field is applied parallel to the plane of the ring.
Noticing that the low energy Hamiltonians for various NRSs have the same generic structure, we employ the continuum model for SrIrO$_3$ as an example of the NRS.
It is explained later that our results are generic and can be applied to other materials as well.
We show that NRS's Landau levels simulate the adiabatic transition from two decoupled graphenes to a Bernal stacked graphene bilayer under the magnetic field with an artificial parallel component as a function of the conserved momentum.
During that process, almost flat Dirac zero modes are found inside the nodal ring.
For some parameters, not far from the realistic ones, the 3D quantum Hall effect (QHE) may occur with a little doping.
Also, we suggest that the nodal ring can be probed by the measurements of density of states (DOS) under the magnetic field.

\begin{figure}
\includegraphics[width=1\columnwidth]{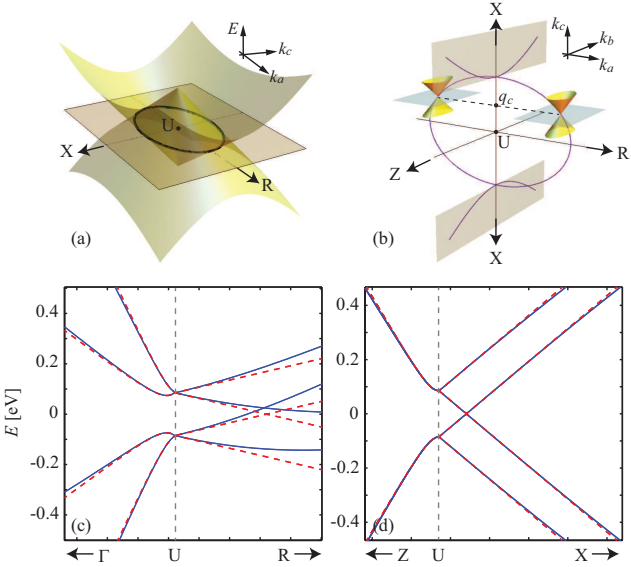}
\caption{(Color online) (a) Low energy band structures on the $k_b=-\pi$ plane where the nodal ring (black ellipse) resides. (b) The nodal chain appears when $2t_2 = t_4$. Two coupled doubly degenerate Dirac cones reside at the points where the nodal lines intersect the plane for given $q_c$. In (c) and (d), we compare energy spectra around U point calculated from the full tight binding Hamiltonian (solid) and the continuum model $\mathcal{H}^\mathrm{U}$ (dashed). }
\label{fig:bands}
\end{figure}

\textit{The model for nodal ring semi-metal.}--- We consider the continuum limit of the tight binding (TB) model for SrIrO$_3$ near the U point ($\v k_\mathrm{U} = (0,-\pi,\pi) $) in the Brillouin zone (Fig. \ref{fig:bands} (a)) \cite{supp}.:
\be
\mathcal{H}^{\mathrm{U}} &=& 2t_0 q_b\tau_z - t_1 q_b\sigma_z\tau_y -t_0 q_c\nu_x +\frac{t_2}{2} q_c (\sigma_x -\sigma_y)\nu_y\tau_z \nn
&& -\frac{1}{2}\left\{ (t_3q_a+t_4q_b)\sigma_x +(t_3q_a+t_4q_b)\sigma_y \right\}\nu_z\tau_y \nn
&& +t_5(\sigma_x -\sigma_y)\nu_x\tau_y
\ee
where $\v q= \v k - \v k_\mathrm{U}$ and $a,b,c$ represent orthorhombic directions of the lattice. 
Here, $\sigma_\alpha$, $\tau_\alpha$ and $\nu_\alpha$ are Pauli matrices where $\sigma_\alpha$ is for the $J_{\mathrm{eff}}=1/2$ Kramers doublet and $\tau_\alpha$($\nu_\alpha$) is for the sublattices $B$ and $R$ ($Y$ and $G$)\cite{chen}.
The realistic TB parameters are known as $t_0=-0.6$, $t_1=-0.15$, $t_2=0.13$, $t_3=-0.2$, $t_4=0.4$ and $t_5=0.06$ in eV.
We consider, however, a wide range of the TB parameters since we are interested in the generic properties of NRSs

$\mathcal{H}^{\mathrm{U}}$ has both the time-reversal and chiral symmetry ($\mathcal{C} = \sigma_z \nu_y \tau_z$).
As a result, we have particle-hole symmetric doubly-degenerate dispersion relations as follows.
\be
E^{\mathrm{U}}_{\zeta,\zeta^\prime}(\v q) &=& \zeta\Big\{ (v_a q_a)^2 +(v_b q_b)^2 +(v_c q_c)^2 + 2t^{2}_5 \\
&& +\zeta^\prime \left( 8t^{2}_5((v_a q_a)^2+(v_c q_c)^2) +v_{d}^4 q_b^2 q_c^2 \right)^{\frac{1}{2}}\Big\}^{\frac{1}{2}} \nonumber \label{eq:band}
\ee
where $v_a =|t_3|/\sqrt{2}$, $v_b = \sqrt{(t_4^2 + 2(4t_0^2 +t_1^2))/2}$, $v_c = \sqrt{(2t_0^2 +t^2_2)/2}$, $v_d = \{ 4(v_bv_c)^2 - 2t_0^2(t_4-2t_2)^2\}^{1/4}$ and $\zeta,\zeta^\prime=\pm 1$.
Near the U point, they are in good agreement with those of the original TB model as compared in Fig. \ref{fig:bands}(c) and (d).
In the continuum model, the nodal ring is just an ellipse satisfying $4 t_5^2 = t_3^2 q_a^2 + (2t_0^2 +t_{2}^{2}) q_c^2$ with major and minor radii given by $r_a = 2|t_5/t_3|$ and $r_c = 2|t_5|/(2t_0^2 +t_2^{2})^{1/2}$.
The ring exists unless $t_3 = 0$ or $t_0=t_2=0$.
Since the dispersion around the ring is linear along the radial and $k_b$ direction while constant along the nodal line, DOS is proportional to energy and vanishing at the Fermi level.
Interestingly, when $ 2t_2=t_4$, we have extra nodal lines for $|q_c|>r_c$ in $k_a = 0$ plane charaterized by a hyperbolic curve $4t_5^2 = (2t_0^2 +t^{2}_2)q_c^2 -2(4t_0^2+t_1^{2}+2t^{2}_2)q_b^2$ as depicted in Fig. \ref{fig:bands} (b).
We call those consecutively occurring nodal lines as the nodal chain.
While those results are valid only around the U point, one can observe the nodal chain structure in the full TB model when it respects the chiral symmetry $\{\mathcal{H}^{\mathrm{TB}},\mathcal{C}\}=0$ which is realized when $t_{xy}=t_d=0$.

\textit{Numerical analysis of Landau level quantization.}---Now, we consider the Landau level quantization of the NRS.
We assume that the magnetic field is applied along $c$-direction so that it is parallel to the plane of the ring.
We neglect the Zeeman splitting ($\sim 10^{-4}B[\mathrm{T}]$eV) \cite{zhang2} since it is much smaller than the Landau level spacings of this system which is order of 10 meV when $B=1$T as will be shown later.
Using the quantization scheme $q_a = (a+a^\dag)/\sqrt{2}l_\mathrm{B}$ and $q_b = i(a-a^\dag )/\sqrt{2}l_\mathrm{B}$, the Landau level wavefunction is expressed in the form $\Psi=\sum_{n=0}^\infty [ c^{B\uparrow}_n, c^{R\uparrow}_n , c^{Y\uparrow}_n , c^{G\uparrow}_n , c^{B\downarrow}_n , c^{R\downarrow}_n , c^{Y\downarrow}_n , c^{G\downarrow}_n ]^{\mathrm{T}}u_n$, where $u_n$ is the simple harmonic oscillator (SHO) eigenfunction \cite{rhim}.
If we separate the eigenvector into two pieces as $\mathbb{A}_n = [ c^{B\uparrow}_n , c^{R\uparrow}_n , c^{B\downarrow}_n , c^{R\downarrow}_n ]^\mathrm{T}$ and $\mathbb{B}_n = [ c^{Y\uparrow}_n , c^{G\uparrow}_n , c^{Y\downarrow}_n , c^{G\downarrow}_n ]^\mathrm{T}$, they satisfy the following coupled secular equations.
\be
\varepsilon\mathbb{A}_n &=& \sqrt{n+1}\v M_{\mathrm{+}}\mathbb{A}_{n+1}+\v L_{\mathrm{+}}\mathbb{B}_n + \sqrt{n}\v M^\dag_{\mathrm{+}}\mathbb{A}_{n-1}, \label{eq:landau1}\\
\varepsilon\mathbb{B}_n &=& \sqrt{n+1}\v M_{\mathrm{-}}\mathbb{B}_{n+1}+\v L_{\mathrm{-}}\mathbb{A}_n + \sqrt{n}\v M^\dag_{\mathrm{-}}\mathbb{B}_{n-1}, \label{eq:landau2}
\ee
where
\be
\v M_\zeta &=& \sum_{m=1}^2\frac{it_{m-1}}{2^{m-\frac{3}{2}}l_\mathrm{B}}\boldsymbol\Gamma_{m5}-\zeta \sum_{n=3}^4\frac{(-1)^m t_3-it_4}{2\sqrt{2}l_\mathrm{B}}\boldsymbol\Gamma_{1n},  \\
\v L_\zeta &=& -t_0 q_c -\zeta\frac{it_2 q_c}{2}(\boldsymbol\Gamma_{45} -\boldsymbol\Gamma_{35}) -t_5 (\boldsymbol\Gamma_{14} -\boldsymbol\Gamma_{13}).~~
\ee
Here, we adopt the representations of the Dirac gamma matrices in Ref. \cite{murakami}.
The coefficient is assumed to be zero when its subscript $n$ is negative.

\begin{figure}
\includegraphics[width=1\columnwidth]{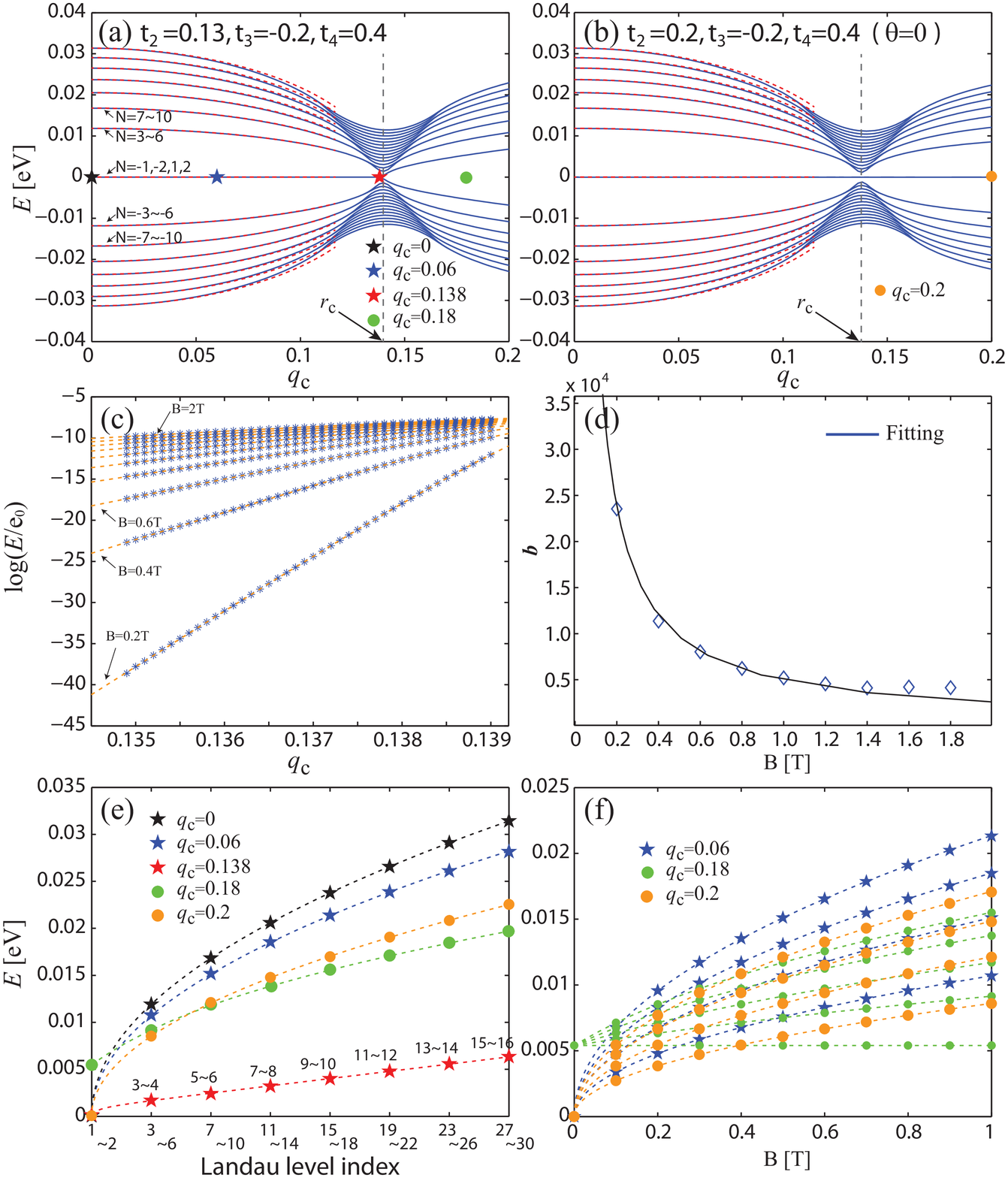}
\caption{(Color online) In (a) and (b), we plot Landau levels of the nodal ring semi-metal  as functions of $q_c$ for different tight binding parameters when the magnetic field (B=1T) is along $c$ direction. Blue solid lines are numerical results while red dashed curves are analytic ones obtained from the fictitious graphene bilayer model near the ring's center. Central Landau levels ($N=1,2$) near the edge of the nodal ring are drawn by markers ($*$) and fitted by a formula $\varepsilon \sim \pm(1-a q_c^2)^\alpha q_c^\beta \exp(b q_c^2)$ for various magnetic fields in (c). Here, $e_0 =1eV$ is introduced to make the argument of the logarithm dimensionless. We show in (d) that the coefficient $b$ in the exponent of the fitting function depends linearly on $B^{-1}$ or $l_\mathrm{B}^{2}$. In bottom panels, we show how the energy depends on the Landau level index and the magnetic field for given $ q_c$. Deep inside the ring ($ q_c=$0, 0.06), we have $\sqrt{NB}$ dependence of the Dirac particle. Those Dirac-like features are lost near the ring's boundary or outside the ring as shown for $q_c=0.138$ and 0.18. In the case of the nodal chain, however, we recover the $\sqrt{NB}$ behavior for larger momenta. }
\label{fig:landau}
\end{figure}

By solving the above equations numerically, we plot the Landau level spectra as functions of the conserved momentum $q_c$ in Fig. \ref{fig:landau}(a) and (b).
Strictly speaking, each band is only doubly degenerate, but one can see almost four-fold degeneracies inside the ring away from the vicinity of its edge at $q_c = r_c$.
We label them by nonzero integer $N$ in such a way that levels with positive(negative) energies are marked by positive(negative) integers in increasing(decreasing) order from central ones near zero energy.

We observe doubly degenerate two bands which are almost flat near zero energy inside the nodal ring ($q_c < r_c$).
Those flat bands are fitted nicely by a formula $\varepsilon \sim \pm(1-a q_c^2)^\alpha q_c^\beta \exp(bq_c^2)$ near the ring's edge as shown in Fig. \ref{fig:landau}(c).
The exponent's coefficient $b$ is found to be approximated as $1.886\times l_\mathrm{B}^{2}$ as plotted in Fig. \ref{fig:landau}(d).
As a result, the energies of central Landau levels ($N=\pm 1,2$) rapidly reduce to zero as we go inside the ring from $q_c = r_c$ by an amount of $\delta q_c \sim 1/(r_c l_\mathrm{B}^2)$.
While the splitting between those flat modes becomes finite outside the ring ($q_c > r_c$), the only exception is when $2t_2 = t_4$ for which we have four zero modes for any value of $q_c$.

In addition, we find that, when $q_c$ is well inside the ring, our Landau level spectra are Dirac-like as shown in Fig. \ref{fig:landau} (e) and (f).
The energies are proportional to the square root of the Landau level index and magnetic field.
On the other hand, the dispersions cannot be fitted by the Dirac Landau levels and show linear behaviors for large Landau level indices and magnetic field if $q_c$ is close to the ring's boundary or outside the ring.
However, when the nodal chain appears for $2t_2=t_4$, one can have Dirac Landau levels again for larger $q_c$.

\textit{Interpretation via Graphene bilayer under a tilted magnetic field}--- To understand the nature of the peculiar Landau level structures of the NRS, we introduce another Hamiltonian $\mathcal{H}^\mathrm{D}$ which is in a block-diagonalized form of two $4\times 4$ submatrices.
Since it gives us exactly the same band structures and Landau levels, it can be regarded as a unitary transform of $\mathcal{H}^\mathrm{U}$.
Those two submatrices of $\mathcal{H}^\mathrm{D}$ are given by
\be
\mathcal{H}^\mathrm{D}_{K(K^\prime)}=\bpm \v h_{K(K^\prime)}^+ & \v g_\theta \\ \v g_\theta^\dag & \v h_{K(K^\prime)}^-\epm,
\ee
where 
\be
&\v h_{K(K^\prime)}^\zeta = v_a(q_a -\zeta p_a)\mu_x \pm v_bq_b\mu_y, & \label{eq:h_in}\\
&\v g_\theta = v_cq_c (\mu_x-i\cos\theta\mu_y-i\sin\theta\mu_z)& \label{eq:g_in}
\ee
for $q_c \leq r_c$ and
\be
&\v h_{K(K^\prime)}^\zeta = v_aq_a\mu_x \pm v_b(q_b-\zeta p_b)\mu_y +\zeta m_0q_c\mu_z, & \label{eq:h_out}\\
&\v g_\theta = \sqrt{2}t_5 \mu_x & \label{eq:g_out}
\ee 
for $q_c > r_c$.
Here, $\mu_\alpha$ is the Pauli matrix and the plus-minus sign denotes $K$($K^\prime$) valley.
$\v h_{K(K^\prime)}^\zeta$ is the Dirac Hamiltonian with the dispersions centered at $\v q =(\zeta p_a,0,q_c)$  for $q_c \leq r_c$ and $\v q =(0,\zeta p_b,q_c)$ for $q_c > r_c$ where $p_a=\sqrt{(2t_5^{2}-(v_c q_c)^2)/v_a^2}$ and $p_b=v_d^2 q_c/2v_b^2$.
When $q_c>r_c$, we have the mass term with $m_0 = \sqrt{v_c^2-v_d^4/4v_b^2} =v_c\sin\theta$ while it is massless inside the ring.
In the mixing term $\v g_\theta$, $\cos^2\theta = v_d^4/4(v_bv_c)^2$.
The transformed Hamiltonian gives us exactly the same energy spectra in Eq. (\ref{eq:band}).

Since $\v h_{K}^\zeta$ and $\v h_{K^\prime}^\zeta$ are the Dirac Hamiltonians at different valleys $K$ and $K^\prime$, one can interpret $\mathcal{H}^\mathrm{D}$ as a Hamiltonian of graphene bilayer with fictitious interlayer coupling $\v g_\theta$.
Here, $\zeta = +$ and $-$ correspond to the upper and lower layer of the artificial graphene bilayer and we denote its basis as $\psi =[c_{A_+}, c_{B_+}, c_{A_-}, c_{B_-} ]^\mathrm{T}$. 
Furthermore, the position of Dirac points in the upper and lower layer are shifted in opposite direction.
For instance, inside the ring ($q_c <r_c$), two Dirac cones are placed at $\v q = (p_a,0,q_c)$ and $(-p_a,0,q_c)$ in the $q_c = const$ plane in the upper and lower layer of the graphene bilayer as shown in Fig. \ref{fig:bands}(b).
This kind of shift can be realized when the magnetic field is applied parallel to graphene layers  with strength $B_\parallel = 2\hbar p_a/ed$ where $d$ is the inter-layer distance \cite{pershoguba1,pershoguba2}.
This fictitious parallel magnetic field has its maximum value at the ring's center ($q_c=0$) and vanishes at the ring's boundary ($q_c=r_c$).
The system transforms from two copies of graphene monolayers to a graphene bilayer with interlayer coupling $\v g_\theta$ as increasing $q_c$ since $\v g_\theta$ is proportional to $q_c$.

Applying the magnetic field along the $c$ direction, we obtain the Landau level dispersions numerically and find that they are exactly the same as the ones from (\ref{eq:landau1}) and (\ref{eq:landau2}).
Now, we discuss detailed properties of the Landau level spectra of the NRS. 

(\textit{i}) Deep inside the ring ($q_c \ll r_c$), the Landau levels of the NRS at each $q_c$ can be considered as those of the four independent anisotropic Dirac particles.
When $q_c =0$, $\v g_\theta$ vanishes and we have four decoupled anisotropic Dirac Hamiltonians.
In this case, the Landau level spectrum is $\varepsilon_m^{\pm} = \pm v_f l_\mathrm{B}^{-1}\sqrt{2m}$ with four-fold degeneracy where $m$ is an integer and $v_f = (v_av_b)^{1/2}$.
To be precise, one has exact zero modes only at $q_c =0$.
However, the Dirac-like feature is maintained up to quite large momenta where $\v g_\theta$ is finite because the mixing between wavefunctions on two Dirac cones is exponentially small ($\sim e^{-p_a^2 l^2_\mathrm{B}}$) with the distance between two Dirac cones, $2p_a$, being maximized at $q_c=0$.
In this regime, the main role of the finite coupling $\v g_\theta$ is the renormalization of the Fermi velocity.
One can project out the contribution of the lower layer and construct an effective low energy Hamiltonian around the Dirac point of the upper layer by using the resolvent $(\varepsilon - \v h^-_{K(K^\prime)})^{-1}$ \cite{mccann,petersen}. 
This gives us a generalized eigenvalue problem of the form $\varepsilon\v (\v I+(2v_a p_0)^{-2})\v g_\theta \v g^\dag_\theta) \psi_+ = (\v h^+_{K(K^\prime)} - (2v_a p_0)^{-2}\v g_\theta \v h^-_{K(K^\prime)} \v g^\dag_\theta)\psi_+$ where $\psi_+$ is the wavefunction for the upper layer.
For an intuitive picture, let us focus on the case $\theta \ll 1$.
Notice that the realistic TB parameters ($t_2=0.13$, $t_3 =-0.2$ and $t_4 =0.4$) correspond to this limit ($\cos\theta = 0.9969$). 
After a transformation to an orthonormal basis set \cite{halpern}, we arrive at a quite simple form of the effective Hamiltonian for the upper layer, $(1+(v_c q_c/v_a p_a)^{2})^{-1/2}\v h^+_{K(K^\prime)}$, with the renormalized Fermi velocity $v_f^\prime = (1+(v_c q_c/v_a p_a)^{2})^{-1/2}v_f$.
In Fig. \ref{fig:landau}(a) and (b), we show by red dashed lines that the Dirac Landau levels with the renormalized Fermi velocity have good agreements with the numerics when $q_c \ll r_c$.

(\textit{ii}) We have, as shown in Fig. \ref{fig:landau}(b), four exact zero modes for arbitrary $q_c$ when $\theta =0$ ($2t_2 = t_4$) where $\v g_\theta$ has only the inter-layer coupling $2v_cq_c$ between $A_-$ and $B_+$ sites.
In this case, our fictitious model is exactly the same as the graphene bilayer with Bernal stacking under a parallel magnetic field.
Although we must rely on the numerics to analyze all Landau levels due to the fictitious in-plane component of the magnetic field, one can show that we have exact zero modes for arbitrary momentum $q_c$.
Here, to avoid unnecessary complexity, we only provide analytic solutions for the case $q_c < r_c$ \cite{supp}.
We define the ladder operator for the upper ($\zeta=+1$) and lower ($\zeta=-1$) layer as $a_\zeta = l_\mathrm{B}(v_a(q_a -\zeta p_a) +iv_b q_b)/\sqrt{2}v_f$ which satisfy $a_+ = a_- -\sqrt{2}v_al_\mathrm{B}p_a/v_f$.
They have their own SHO eigenfunctions $u_n^-$ and $u_n^+$ which are shifted spatially from each other in $b$-direction. 
In this case, one can find four zero energy eigenvectors as $\psi^0_{K,1} = [ u^+_0 , 0 , 0 , 0]^\mathrm{T}$ and $\psi^0_{K,2} = c_0[u^-_0 , 0 , \gamma_0 u^-_0 , 0]^\mathrm{T}$ for $K$ valley and $\psi^0_{K^\prime,1} = [ 0 , 0 , 0 , u^-_0]^\mathrm{T}$ and $\psi^0_{K^\prime,2} = c_0[ 0 , -\gamma_0 u^+_0 , 0 , u^+_0]^\mathrm{T}$ for $K^\prime$ valley, where $\gamma_0 = p_a v_a/(v_c q_c)$ and $c_0=(1+\gamma_0^2)^{-1/2}$ is the normalization factor.
One can also find analytic form of the eigenfunctions at the zero energy for $q_c \geq r_c$ in the similar way.
The reason why we still have zero modes outside the ring is that we have four massless Dirac cones along the hyperbolic nodal line in $k_a=0$ plane (Fig. \ref{fig:bands}(b)) since $m_0=0$ for $\theta =0$ \cite{footnote}.

(\textit{iii}) On the other hand, the four-fold symmetry of the zero modes is broken for nonzero $\theta$ although the splitting between them is almost negligible. 
By using the zero energy solutions inside the ring, we estimate their splitting for finite $\theta$ near $q_c =0$ as $\Delta \approx 2|\langle \psi^0_{K(K^\prime),1}| [\mathcal{H}^\mathrm{D}_{H(K^\prime)}(\theta) -\mathcal{H}^\mathrm{D}_{K(K^\prime)}(0)]| \psi^0_{K(K^\prime),2} \rangle| = d_0 q_c (1-a_0q_c^2)^{1/2} \exp(-b_0q_c^2)$ where $a_0 = v_c^2/(2t_5^{2})$, $b_0=v_c^2 l^2_\mathrm{B}/v_a^2$ and $d_0 = v_c\exp(-2t_5^{2}l^2_\mathrm{B}/v_a^2)$.
The splitting of the Lowest Landau level is extremely small near $q_c =0$ and looks almost flat due to the factor $l^2_\mathrm{B}$ in the exponent of $d_0$ which reflects the effect of the huge fictitious parallel magnetic field near $q_c=0$.
The derivation of the above is only valid around $q_c =0$ where the mixing from higher Landau levels is minimal so that the projection onto the central Landau levels is safe.
However, inspired by this formula, we could obtain the fitting function for the bands near the ring's edge as shown in Fig. \ref{fig:landau}(c). 

(\textit{iv}) At $q_c =r_c$ and $\theta=0$, where $B_{\parallel}=0$, the situation becomes the usual Bernal stacked graphene bilayer under the perpendicular magnetic field and the Landau levels are evaluated as $\varepsilon_n^{\zeta,\zeta^\prime}= \zeta^\prime 2^{1/2}v_c q_c ( \lambda_{1,n} +\zeta ( \lambda_{2,n} +\omega^4)^{1/2})^{1/2}$
where $\lambda_{m,n}=1+m(2n-1)\omega^2$ and $\omega^2 = v_f^2/(2v_c^2 q_c^2 l_\mathrm{B}^2)$.
The Lanadu levels near the Fermi level are described by $\varepsilon_n^{-,\xi}$.
In the low energy regime, it is approximated to $\varepsilon_n^{-,\xi}\approx \xi v_f^2/(v_c q_c l_\mathrm{B}^2)\sqrt{n(n-1)}$ which is the well-known Landau levels of graphene bilayer \cite{mccann}.

\textit{3D quantum Hall effect}--- One of the interesting results for $\theta =0$ case (Fig. \ref{fig:landau}(b)) is that the zero energy flat modes are separated from other bands with a finite gap.
In this case, one can have the 3D QHE by a slight doping as indicated by Halperin \cite{halperin}.
He showed that the conductivity tensor should be in the form of $\sigma_{ij} = e^2/(2\pi h)\epsilon_{ijk}G_{k}$ where $\epsilon_{ijk}$ is the Levi-Civita symbol and $\v G$ is the reciprocal vector \cite{halperin,bernevig}.
When $q_c=0$ and $q_c=r_c$, the Hall conductance in the $ab$ plane is given by $\sigma^{2\mathrm{D}}_{ab} = 2e^2/h$ since our model is equivalent to the decoupled spinless graphene layers and Bernal stacked graphene bilayer respectively.
As we do not have any gap closing, the Hall conductances at other momenta are also $\sigma^{2\mathrm{D}}_{ab} = 2e^2/h$ due to the adiabatic continuation.
Then, the three dimensional Hall conductance is evaluated as 
\be
\sigma_{ab} = \int \frac{d q_c}{2\pi}\sigma^{2\mathrm{D}}_{ab} = \frac{e^2}{2\pi h} \frac{4\pi}{a_c} 
\ee
where $a_c$ is the lattice constant along the $c$ direction.
In the real material, it is expected that $\theta \neq 0$ and also there may be other electron or hole pockets at the Fermi level.
In this case, we expect that the Hall conductance would be order of $\sim e^2/ha_c$ even though it is not strictly quantized.
Further, we expect the strain effect may be used to tune the TB parameters close to the ideal case of the above.

\textit{Discussion}--- Although we used a continuum model for SrIrO$_3$, the flat 3D Landau levels at the Fermi energy can be found in any semi-metals with nodal ring dispersion, where the energy spectra are linear along the perpendicular directions of the nodal line so that one can find Dirac cones for given momentum parallel to the ring's plane as shown in Fig. \ref{fig:bands}(b).
Recently, other candidate materials for the NRSs with the above properties have been suggested, such as Cu$_3$NZn, Cu$_3$NPd \cite{kim,zeng2}, Ca$_3$P$_2$ \cite{xie}, LaN \cite{zeng} and so on \cite{xu2,zeng1,zeng2}.
For example, the low energy Hamiltonian of Cu$_3$NZn, for a given $q_y$ ($\v q_\perp = (q_x,q_y)$), is given by $H \sim 2b_\perp q_{0}(q_x-q_{0})\tau_z + vq_r\tau_y$ where $q_0 = \pm(-\Delta\epsilon/b_\perp -q_y^2)^{1/2}$ are the positions of two Dirac cones.
It is also noticed that Ca$_3$P$_2$ has similar electronic structures from the band crossings along $M \Gamma K$ and the linear DOS around the nodal ring.
Among these, Ca$_3$P$_2$ may be the most promising since its ring is free from other electron or hole pockets and has a sizable radius.
Flat Dirac Landau levels would appear in these systems when the magnetic field is applied in the direction of the ring's plane.
Because the flat bands have prominent peak in the DOS, one might identify the existence of the nodal ring by the scanning tunneling microscopy even if it is buried in other dispersive bands.

This work was supported by the NSERC of Canada, the CIFAR, and the Center for Quantum Materials at the University of Toronto. We thank Y. Chen, H.- Y. Kee, H.-S. Kim and J. Kim for useful discussions.



\begin{references}

\bibitem{novoselov} K. S. Novoselov, A. K. Geim, S. V. Morozov, D. Jiang, M. I. Katsnelson, I. V. Grigorieva, S. V. Dubonos, and A. A. Firsov, Nature (London) \textbf{438}, 197 (2005).

\bibitem{zhang1} Y. Zhang, Y.-W. Tan, H. L. Stormer, and P. Kim, Nature (London) \textbf{438}, 201 (2005).

\bibitem{castro} A. H. Castro Neto, F. Guinea, N. M. R. Peres, K. S. Novoselov, and A. K. Geim
Rev. Mod. Phys. \textbf{81}, 109 (2009).

\bibitem{mccann} Edward McCann and Vladimir I. Fal’ko, 
Phys. Rev. Lett. \textbf{96}, 086805 (2006).

\bibitem{yang} Kai-Yu Yang, Yuan-Ming Lu, and Ying Ran,
Phys. Rev. B \textbf{84}, 075129 (2011). 

\bibitem{yang2} Bohm-Jung Yang, Yong Baek Kim, Phys. Rev. B \textbf{82}, 085111 (2010).

\bibitem{krempa} William Witczak-Krempa, Yong Baek Kim, Phys. Rev. B \textbf{85}, 045124 (2012).

\bibitem{moon} E.-G. Moon, C. Xu, Y. B. Kim, and L. Balents, Phys. Rev. Lett. \textbf{111}, 206401 (2013).

\bibitem{qi} S. Raghu, Xiao-Liang Qi, C. Honerkamp, and Shou-Cheng Zhang,
Phys. Rev. Lett. \textbf{100}, 156401 (2008). 

\bibitem{rhim} Jun-Won Rhim and Yong Baek Kim, 
Phys. Rev. B \textbf{91}, 115124 (2015).


\bibitem{burkov} A. A. Burkov and Leon Balents,
Phys. Rev. Lett. \textbf{107}, 127205 (2011). 

\bibitem{wan} Xiangang Wan, Ari M. Turner, Ashvin Vishwanath, and Sergey Y. Savrasov,
Phys. Rev. B \textbf{83}, 205101 (2011). 


\bibitem{xu} Gang Xu, Hongming Weng, Zhijun Wang, Xi Dai, and Zhong Fang,
Phys. Rev. Lett. \textbf{107}, 186806 (2011). 

\bibitem{young} S. M. Young, S. Zaheer, J. C. Y. Teo, C. L. Kane, E. J. Mele, and A. M. Rappe,
Phys. Rev. Lett. \textbf{108}, 140405 (2012). 

\bibitem{wang} Zhijun Wang, Yan Sun, Xing-Qiu Chen, Cesare Franchini, Gang Xu, Hongming Weng, Xi Dai, and Zhong Fang, Phys. Rev. B \textbf{85}, 195320 (2012). 


\bibitem{hasan} M. Z. Hasan and C. L. Kane, Rev. Mod. Phys. \textbf{82}, 3045 (2010); X.-L. Qi and S.-C. Zhang, ibid. \textbf{83}, 1057 (2011).

\bibitem{matsuura} Shunji Matsuura, Po-Yao Chang, Andreas P. Schnyder and Shinsei Ryu, New J. Phys. \textbf{15}, 065001 (2013).


\bibitem{burkov} A. A. Burkov, M. D. Hook, and Leon Balents, Phys. Rev. \textbf{B} 84, 235126 (2011)

\bibitem{carter} Jean-Michel Carter, V. Vijay Shankar, M. Ahsan Zeb, and Hae-Young Kee, 
Phys. Rev. B \textbf{85}, 115105 (2012).

\bibitem{chen} Yige Chen, Yuan-Ming Lu, and Hae-Young Kee, 
Nature Communications \textbf{6}, 6593 (2015).

\bibitem{kim} Youngkuk Kim, Benjamin J. Wieder, C. L. Kane, and Andrew M. Rappe, arXiv:1504.03807.

\bibitem{xie} Lilia S. Xie, Leslie M. Schoop, Elizabeth M. Seibel,
Quinn D. Gibson, Weiwei Xie, and Robert J. Cava, arXiv:1504.01731.

\bibitem{zeng} Minggang Zeng, Chen Fang, Guoqing Chang, Yu-An Chen,
Timothy Hsieh, Arun Bansil, Hsin Lin, and Liang Fu, arXiv:1504.03492.

\bibitem{xu2} Su-Yang Xu, Nasser Alidoust, Ilya Belopolski, Chenglong Zhang, Guang Bian, Tay-Rong Chang, Hao Zheng, Vladimir Strokov, Daniel S. Sanchez, Guoqing Chang, Zhujun Yuan, Daixiang Mou, Yun Wu, Lunan Huang, Chi-Cheng Lee, Shin-Ming Huang, BaoKai Wang, Arun Bansil, Horng-Tay Jeng, Titus Neupert, Adam Kaminski, Hsin Lin, Shuang Jia, M. Zahid Hasan, arXiv:1504.01350.

\bibitem{zeng1} Hongming Weng, Yunye Liang, Qiunan Xu, Yu Rui, Zhong Fang, Xi Dai, Yoshiyuki Kawazoe, arXiv:1411.2175.

\bibitem{zeng2} Rui Yu, Hongming Weng, Zhong Fang, Xi Dai, Xiao Hu, arXiv:1504.04577.

\bibitem{zeng3} Hongming Weng, Chen Fang, Zhong Fang, B. Andrei Bernevig, and Xi Dai, Phys. Rev. X \textbf{5}, 011029.




\bibitem{supp} See Supplemental Material for the introduction of the full tight binding model of SrIrO$_3$ and details for the analysis of the bilayer graphene-like model $\mathcal{H}^\mathrm{D}$.

\bibitem{zhang2} Lunyong Zhang, Y. B. Chen, Binbin Zhang, Jian Zhou, Shantao Zhang, Zhengbin Gu, Shuhua Yao, and YanFeng Chen, J. Phys. Soc. Jpn. \textbf{83}, 054707 (2014).

\bibitem{murakami} Shuichi Murakami, Naoto Nagaosa, and Shou-Cheng Zhang, 
Phys. Rev. B \textbf{69}, 235206 (2004).


\bibitem{pershoguba1} Sergey S. Pershoguba and Victor M. Yakovenko
Phys. Rev. B \textbf{82}, 205408 (2010).

\bibitem{pershoguba2} Sergey S. Pershoguba, D. S. L. Abergel, Victor M. Yakovenko, and A. V. Balatsky
Phys. Rev. B \textbf{91}, 085418 (2015).


\bibitem{petersen} L. Petersen and P. Hedegård, Surf. Sci. \textbf{459}, 49 (2000).

\bibitem{halpern} V. Halpern 1971 J. Phys. C: Solid State Phys. \textbf{4} L369 (1971).


\bibitem{footnote} Unlike the case $q_c<r_c$, when $q_c>r_c$, the artificial parallel magnetic field cannot keep the flatness of the zero modes at $\theta=0$ due to the mass terms in each Dirac Hamiltonian. As a result, we have large splittings between central Landau levels as shown in Fig. \ref{fig:landau}(a) and (b) even if $\theta$ is small.



\bibitem{halperin} B. I. Halperin, Jpn. J. Appl. Phys. Suppl. \textbf{26}, 1913 (1987).

\bibitem{bernevig} B. Andrei Bernevig, Taylor L. Hughes, Srinivas Raghu, and Daniel P. Arovas, Phys. Rev. Lett. \textbf{99}, 146804 (2007)

 
\end{references}
\end{document}